\documentstyle[preprint,aps]{revtex}
\begin{document}


\draft

\title{Theoretical Study of One-dimensional Chains of Metal Atoms in Nanotubes}

\author{Angel Rubio\cite{add2}, Yoshiyuki Miyamoto\cite{add1},
 X. Blase, Marvin L. Cohen, and Steven G. Louie}
\address{
Department of Physics, University of California at Berkeley,
Berkeley, California 94720
and
Materials Sciences Division, Lawrence Berkeley Laboratory,
Berkeley, California 94720
}

\date{\today}

\maketitle
\begin{abstract}


Using first-principles total-energy pseudopotential calculations, we have
studied the properties of chains of potassium and aluminum in nanotubes.
For BN tubes, there is little interaction between the
metal chains and the tubes, and
the conductivity of these tubes is
through carriers located at the inner part of the tube.
In contrast, for small radius
carbon nanotubes, there are two types of interactions: charge-transfer
(dominant for alkali atoms) leading to strong ionic cohesion, and
hybridization (for multivalent metal atoms) resulting in a smaller cohesion.
For Al-atomic chains in carbon tubes, we show that both effects contribute. New
electronic properties related to these confined atomic chains of metal are
analyzed.
\end{abstract}

\pacs{PACS numbers: 71.25.Tn, 36.20.Kd}

\narrowtext

Quasi-one-dimensional metals and conducting nanowires are of great interest
both
from a fundamental point of view and for possible applications. Here we explore
theoretical models for examining the structural and transport
properties of these systems~\cite{Garcia}. Tubule forms of graphitic
carbon~\cite{Iijima} with
their expected interesting electronic and mechanical properties
\cite{Hamada,prl,Ajayan,Dujardin,Guerret,Heer} are ideal candidates as hosts
for one dimensional metal systems. Carbon tubes have been made which are
microns in length~\cite{Iijima,Guerret,Heer,Ajayan1,Iijima1} and have diameters
ranging from larger than 100$\AA $ for multi-wall tubes down to less
than 10$\AA $ for single-wall tubes \cite{Iijima1,Bethune}. Important
effects in the conductivity as a function of length and diameter of
encapsulated metallic nanowires are expected and confinement might induce new
metallic phases.  For small diameter tubes, the captured metal atoms inside
can form an atomic linear-chain~\cite{prl}.
This would provide a new means for producing
ideal one-dimensional metallic chains. The metal capillarity and doping of
tubes
also
raise the possibility of changing the electronic structure of the tube
itself by either charge-transfer or hybridization. Just as in the case of
alkali
doped C$_{60}$, we expect curvature effects to alter electron-phonon and
superconducting
pairing interactions from those of intercalated graphite~\cite{Lorin}.

At present, we know of no evidence of alkali intercalation into nanotubes
experimentally which is in contrast to the results for graphitic intercalation
compounds (GIC's)~\cite{GIC}. Most of the experiments on metal-atom
intercalation have been done on multi-wall tubes with large diameter sizes
(more than 100~$\AA$ diameter)~\cite{Ajayan,Dujardin,Guerret,Ajayan1,Tanaka}.
Transition metals inside the tubes were studied~\cite{Guerret}, and it was
found that the formation of continuous nanowires was connected with the
existence of an incomplete electronic shell in the most stable ion
configuration of the metal. However, in most cases, the filling material was a
crystalline metallic carbide.
Lead, bismuth and manganese incorporation has also been reported
\cite{Ajayan,Ajayan1}.  A more detailed
study~\cite{Dujardin} has proposed surface tension of the liquid phase of the
metals as a key factor in determining whether capillary action (wetting)
occurs. The experimental surface tension threshold found for wetting
is near 190mN/m. Thus, in this picture, fillings using metals with larger
surface tension require external pressure to be experimentally realizable, and
it is expected that typical metals will not be drawn into tube cavities by
capillary action. On the other hand, a strong ionic cohesion for K inside
subnanometer-size carbon tube was theoretically predicted~\cite{prl}. The
assumed classical theory of wetting was therefore concluded to
be inappropriate for this tube size, and incorporation of other
metal-atoms should be possible.

Based on the similarities among graphite and hexagonal boron-nitride compounds,
we have predicted that BN and B$_x$C$_y$N$_z$ will form stable
tubes~\cite{Angel,bcn}.
This was recently proven experimentally by electric arc-discharge
synthesis~\cite{bn_exp} as well as by laser-driven gas-phase chemical reaction
synthesis~\cite{Willaime}. The electronic properties of this new class of
nanotubes are quite different from their carbon counterpart. Namely, the BN
nanotubes
are stable wide band-gap  semiconductors (E$_{\rm g}~\sim$~5.5~eV) independent
of helicity, diameter of the tube, or whether the tube is single
wall or multi-wall. Furthermore, the bottom of the conduction band is a nearly
free electron (NFE) like-state that derives its character from the
weakely bound states of a BN sheet in a band-folding picture~\cite{bn_hex}.
Considering
that insulators are much less polarizable than metals and semimetals, it is
expected that the potential experienced by an electron in the internal vacuum
region close to the BN tube surface will be less binding than that for
graphite. Therefore, we expect BN to behave like an ideal {\it non-interacting}
host for the metal atoms inside. We note here that other composition B-C-N
tubes have also been proposed (BC$_2$N and BC$_3$) as stable~\cite{bcn} and
observed experimentally~\cite{bcn_exp}.
These systems have potential technological applications
and have very interesting electronic properties which can be generally
explained by rolling the corresponding planar sheets. However,
we expect that these B-C-N tubes would not serve as non-interacting hosts
since their band-gaps are small.

The purpose of the present study is to examine two particular cases of
intercalation with metal atoms of
different chemical valence: K and Al atoms. We compare and
contrast the results for BN tubes with those for carbon
nanotubes. The known alkali GIC's~\cite{GIC} and the intercalation of K-atoms
in
hexagonal BN~\cite{Doll} supports the possibility of incorporation of this
metal into small radius nanotubes.
However, there are no reports for Al GIC's and Al
intercalation into hexagonal BN to our knowledge, so no comparison can be
made. We show that BN tubes behave as {\it non-interacting} confining
hosts for metallic chains made of K (or other alkali atoms) and Al. This is in
contrast with metal atoms inside carbon nanotubes where
charge-transfer and hybridization effects are obtained.
In all cases, we expect that reactions in the gas phase would favor the
intercalation of these metallic atoms in both carbon and BN nanotubes.
We conclude that the thin tubes react with metals similarly to the
process of intercalation of the planar graphitic sheets.

Total-energy pseudopotential band-structure calculations are done for the metal
and tube systems within the framework of the local density
approximation (LDA) using a planewave basis set with a cutoff energy of 36 Ry.
The Kleinman-Bylander {\it ab-initio} pseudopotential scheme~\cite{KB}
including core-corrections for the exchange-correlation energy~\cite{Steve}
is used. The core correction is necessary to reproduce the
structural properties of bulk K. The calculations were performed in a
supercell geometry with a 5.5$\AA$ distance between the walls of neighboring
tubes. This distance is large
enough to ensure that tube-tube interactions are negligible.
We use two {\bf k}-points in the one dimensional irreducible Brillouin zone
to get well converged total energies and electronic states.

We first perform a structural minimization for the free standing linear chain
of K and Al atoms. The obtained bond length of the K chain is 4.08~$\AA$
and the Al chain 2.38~$\AA$, as compared to the calculated values of 4.38 $\AA$
for bulk bcc K and 2.84~$\AA$ for bulk fcc Al. The smaller bond-length can be
understood in terms of a reduction of the coordination number going from the
bulk system to the linear chain~\cite{Pauling}. In the case of a K chain, a
small energy lowering (below 0.5~K)~\cite{prl}
was found by dimerization of the K atoms. In contrast,
an energy lowering of the Al chain from dimerization
was not found in the present calculations.
These results however do not rule out a Peierls transition for K and
Al chains. Since the energy gain is extremely small, the calculations require
a large number of sampling points in momentum space.
According to these results, we believe that Peierls distortion would not be
important in practical situations. We have estimated the cohesive energy to
be 0.47~eV/atom and 1.23~eV/atom for the K and Al chains respectively, as
compared to the experimental values of 0.94~eV/atom
and 3.34~eV/atom for bulk K and Al
respectively~\cite{Handbook}. For the Al chain, the large difference in
cohesive
energy with respect to the bulk metal can be related to the incomplete
formation of bulk $sp$-delocalized bands in the one dimensional chain.
Similar effects are expected to happen for the electronic structure when the Al
chain is incorporated into a BN nanotube.

In order to reduce the computational effort, we have assumed that the linear
chain and the nanotube constitute commensurate phases because the change in the
binding energy of the linear chain in going from the theoretical bond length to
the well-matched tube lattice constant is small~\cite{comment1}.
Thus we set the metal chain
bond length to be the same as the periodic distance along the tube axis in
all the calculations reported here. We note that the K chain is well suited
to be incorporated in $(n,0)$ single unit cell tubes or double unit cell
$(n,n)$ nanotubes, while the Al chain fits well in the $(n,n)$ single unit cell
nanotubes~\cite{comment2}. The index notation for the tubes used here is the
same as that given in Ref.~\cite{Hamada}.  Furthermore,
the metal atoms are assumed to be linearly aligned at the tube center. This
is not a severe approximation since for the diameter of the tubes studied, the
metallic linear chain is in the only possible arrangement because of the
geometrical restrictions for the covalent and metallic radii
of the atoms~\cite{Pauling}.

In Ref.~\cite{prl} there is a discussion of
the important role of having small diameter carbon
tubes (diameters close to the GIC's interlayer distance) for maximizing the
heat
of formation for incorporating K atoms ($\sim $1~eV/K-atom).
The heat of formation is obtained by
subtracting the total energy of the doped tube from the sum of the total
energies of separated systems of bulk metal and a pure tube. This
subnanometer diameter tubes do not obey the classical picture of wetting and
capillarity. We here note that non-intercalation of Al atoms
is expected for very large diameter tubes because liquid Al has a
much larger surface tension~\cite{Handbook} than the threshold for
wetting~\cite{Dujardin}. For the intercalation of Al in a C(4,4)-tube, we
find that the heat of formation is negative, in contrast to K.
However, there is a gain in energy of 0.14~eV/atom comparing the energy of
a tube with a linear chain inside to that with the chain separated. We thus
expect intercalation of Al in the gas phase could be experimentally
realizable under pressure for carbon tube bundles.
The main reason for the negative heat of formation
is that the coordination number
for Al is not large enough to form the $sp$-metallic band of bulk Al. Hence we
expect that Al is likely to form larger diameter nanowires
(with a tendency to fcc coordination) when the tube
diameter is larger but still in the nanometer regime. In Fig.~(1) the band
structure of Al inside a C(4,4) nanotube is compared with that of an undoped
tube. In Fig.~(2), a band structure of a free standing Al chain is shown for
the comparison. Upon Al incorporation, the Fermi level is shifted upward
corresponding to one electron transfer per metal atom as in the case
of K-intercalation. On the other hand, some hybridization between the Al-$s$
states with the tube wall is observed. This Al-state is indicated in
Fig.~(1). The Al p$_{z}$-states and p$_{xy}$-states
are above the Fermi-level and also hybridize with the conduction tube states.

To gain insight into the physics of intercalation of K atoms into BN tubes,
we first performed some calculations for the hexagonal planar phase.
Hexagonal BN is
an indirect wide-gap semiconductor with a NFE
state which lies close to the bottom of the conduction band~\cite{bn_hex}. This
NFE state is the one that interacts the most with the K $s$ states. The
position of the K-derived level, around 3~eV above the BN valence band top, is
consistent with photoluminescence experiments~\cite{Doll}. We remark that
the band structure of a single BN-sheet is also a wide band gap insulator but
the NFE state now becomes the bottom of the conduction band. According to our
calculation, the wetting of the planar BN sheet by a K-monolayer is an
exothermic reaction ($\sim$~0.5~eV/K-atom) with a 2$\times$2 unit cell.
In a manner similar to the adsorption of K layers
on graphite~\cite{Kmono}, multiple-layers are
not expected to be adsorbed on the BN sheet. These results support the possible
intercalation of K atoms in small radius BN nanotubes.
In the three cases studied here:
K atoms inside BN(4,4) and BN(8,0), and Al atoms in BN(4,4) the difference in
the binding energy of the two systems (energy of the tube with metal
outside minus the one with metal inside)
is positive but smaller than the value for Al
in carbon nanotubes mainly due to the negligible charge-transfer
from the chain to the BN-wall.
This is an indication of weak interaction between the
metal-chain and the BN-tube. Intercalation however could still
be possible while growing the nanotube in a vapor phase of the metallic
element.

The band structures of K and Al chains in BN(4,4) tubes are depicted
in Fig.~(3). The NFE state interacts with the K chain $s$ states, keeping the
wavefunction of the occupied carrier states
located in the interior of the tube on the K chain (see a contour map of
charge density for the state $\alpha$ at the bottom of Fig.~(3)).
This is different from K in carbon nanotubes, in which the K chain $s$ states
and the carbon NFE state also interact but the corresponding
wavefunctions are unoccupied and the K electrons are donated to the
graphitic wall.
We check that these results are independent of the
diameter and helicity of the BN tube by doing the
calculation for K in BN(8,0). The same picture is obtained with
hybridization among the tube NFE-band and the K-$s$ derived band
forming occupied states with wavefunction inside the tube. The
band structures of the multiwall BN-nanotubes do not change
significantly from those of the single wall tubes, hence we expect
the intercalation to be the same as for the single wall tubes.

The band structure for an  Al chain in a BN(4,4) tube can be
understood by the direct addition of the two band structures
of the isolated systems (BN tube plus free Al chain).
Electronic states in the gap at X are clearly
seen as being derived from Al $s$ and $p_{z}$ linear chain states,
(see Fig.~(2)). This is different from the
carbon nanotubes where the Al $p_{z}$ states are located just close to
the Fermi level.  The Al $s$ state at X is displayed in the bottom part of
Fig.~(3). Also, plotted is one of the states at $\Gamma$
derived from the interaction of the NFE-state with the Al $p_{xy}$
states ($\beta_1$ and $\beta_2$ states in the figure
which have very similar charge density profiles).
We note that all of these states
have charge density concentrated in the inner part of the tube.
Therefore, the conductivity of these doped-tubes will be controlled by carriers
in the inside of the tubes whereas,
in carbon nanotubes, carriers on the tube wall dominate.
When considering the rigid sliding motion of the inner metal chain, we find
that
the change in energy as a function of position is small and comparable to the
one given for K in carbon nanotubes~\cite{prl}.

In conclusion, we have shown that boron-nitride nanotubes may be good hosts for
the incorporation of metal nanowires. This together with the elastic
properties and high thermal conductivity of the BN-matrix suggest that
these systems may have technological applications. Similar studies on
carbon nanotubes show both charge-transfer and hybridization
mechanisms are possible for the formation of the metal/tube system.
The conductivity of the intercalated tubes will be
dominated by carriers in the inner region of the tube for boron-nitride tubes
and on the tube wall for carbon nanotubes. The transition between
macroscopic behavior controlled by surface tension and the
microscopic level incorporation studied here is still an open question.
Noble-metal atoms such as Ag and Au could also be incorporated in these tubes
forming continuous nanowires.
Quantized conductance versus localization effect as a function of length
has been reported for thin Au-wires~\cite{Garcia}. Similar or more striking
effects may be observed in a one-dimensional metal chain inside a tube.

{\em Acknowledgements: }
 This work was supported by National Science Foundation Grant No. DMR-9520554
 and by the Director Office of Energy Research, Office of Basic Energy
 Sciences, Materials Sciences Division of the U.S. Department of Energy under
 Contact No. DE-AC03-76SF00098. The computations were done
 on the CRAY-C90 computer at the San Diego Supercomputer Center.
 X.B. gratefully acknowledges support from the France-Berkeley fund.

\begin{figure}
\caption{ Calculated band structures:  {\bf (a)} (4,4) carbon
 tube and {\bf (b)} Al-intercalated (4,4) carbon tube.
 The location of the Fermi level $E_{F}$ as well as the
 Al-chain derived states are indicated.}
\end{figure}

\begin{figure}
\caption{Band structure of a free standing Al linear chain.}
\end{figure}

\begin{figure}
\caption{ Top, calculated band structures: {\bf (a)} K-intercalated BN(4,4)
 tube, pure BN(4,4) tube in a {\bf (b)} single and {\bf (c)} double
 unit cell, and {\bf (d)} Al-intercalated BN(4,4) tube. We include {\bf (b)}
 and {\bf (c)} (related by folding) to allow comparison with the corresponding
 doped tubes.  We note that the Brillouin zone of the
 K case is half that of Al.  The maxima of the valence
 bands of the undoped tubes is set as the energy origin.
 The location of the Fermi level $E_{F}$ is indicated.
 The NFE state and the states that derive from its interaction with the
 metallic-chain states are indicated and plotted in the bottom panel
 (states  $\alpha$, $\beta_1$, similarly for $\beta_2$).
 Also the Al $s$- and $p_{z}$-states are indicated
 and the Al $s$ charge density is plotted.}
\end{figure}


\begin{references}
\bibitem{comment1} This change in binding is of
 0.011~eV/atom and 0.04~eV/atom for K- and Al-chains, respectively.
\bibitem{comment2} The lattice constant for $(n,0)$ tubes
 is 4.29~$\AA$ and 4.32~$\AA$ for carbon and BN, respectively.
 Meanwhile, the lattice constant for $(n,n)$ tubes is
 2.46~$\AA$ for C and 2.48~$\AA$ for BN.
\end{references}
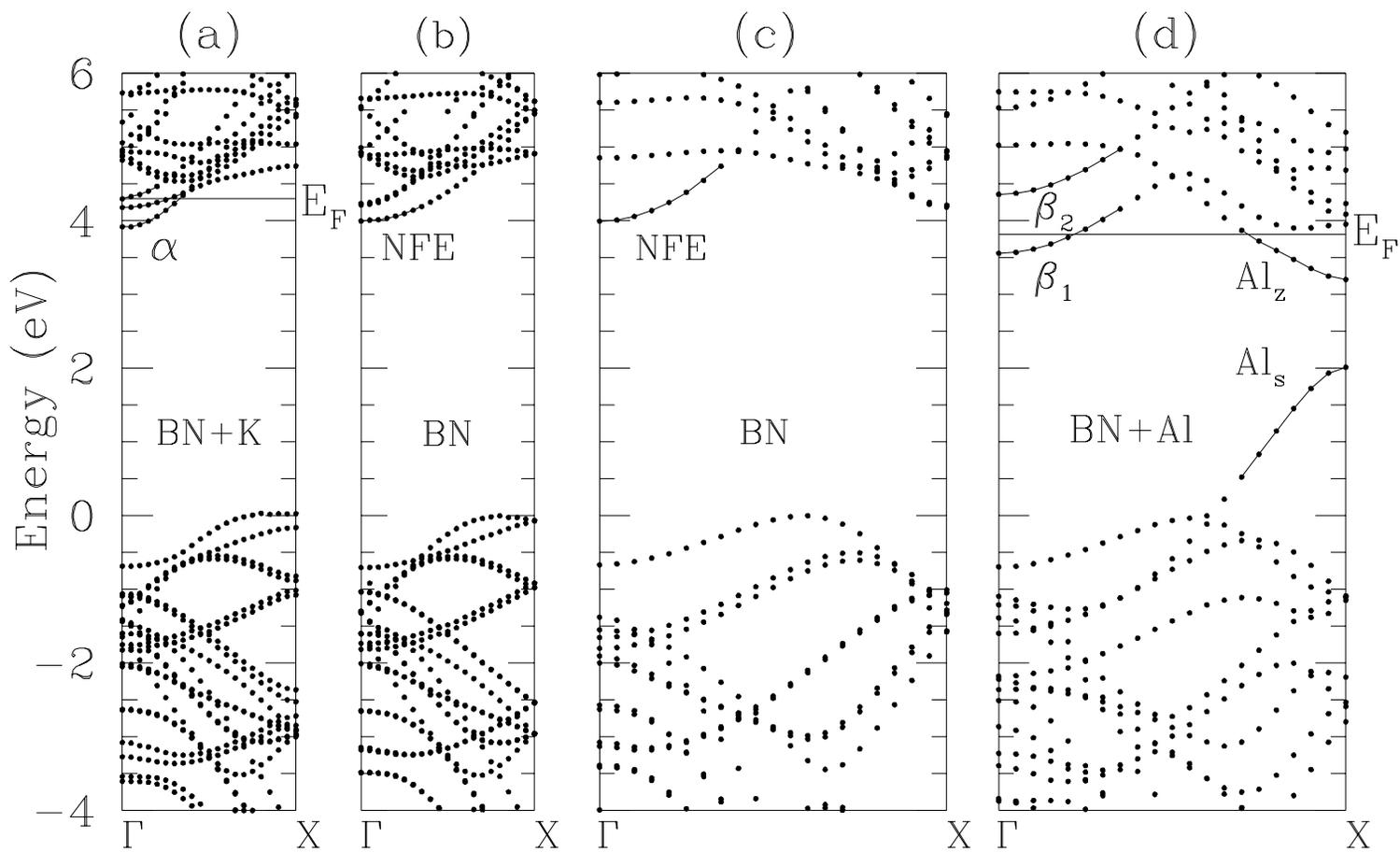
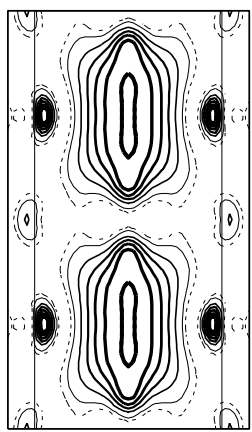
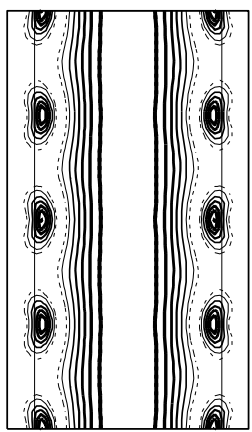
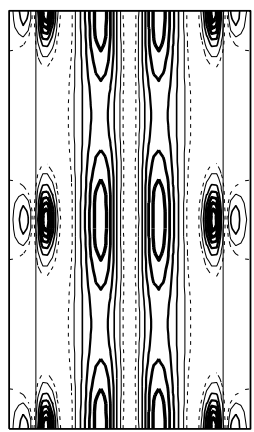
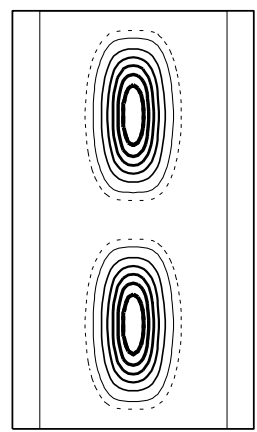
\end{document}